# Mass production of ultra-pure NaI powder for COSINE-200

**KeonAh Shin[1*], JunSeok Choe[1], Olga Gileva[1], Alain Iltis[2], Yena Kim[1], Yeongduk Kim[1,3], Cheolho Lee[1], Eunkyung Lee[1], and HyunSu Lee[1,3*], Moo Hyun Lee[1,3]**

[1]Center for Underground Physics, Institute for Basic Science (IBS), Daejeon 34126, Korea

[2]Damavan Imaging, Troyes, 10430, France

[3]IBS school, University of Science and Technology (UST), Daejeon 34113, Korea

**Correspondence:**
KeonAh Shin and HyunSu Lee
kashin@ibs.re.kr, hyunsulee@ibs.re.kr

**Abstract**

COSINE-200 is the next phase experiment of the ongoing COSINE-100 that aims to unambiguously verify the annual modulation signals observed by the DAMA experiment and to reach the world competitive sensitivity on the low-mass dark matter search. To achieve the physics goal of the COSINE-200, the successful production of the low-background NaI(Tl) detectors is crucial and it must begin from the mass production of the ultra-low background NaI powder. A clean facility for mass-producing the pure-NaI powder has been constructed at the Center for Underground Physics (CUP) in Korea. Two years of operation determined efficient parameters of the mass purification and provided a total of 480 kg of the ultra-pure NaI powder in hand. The potassium concentration in the produced powders varied from 5.4 to 11 ppb, and the maximum production capacity of 35 kg per two weeks was achieved. Here, we report our operational practice with the mass purification of the NaI powder, which includes raw powder purification, recycling of the mother solution, and recovery of NaI from the residual melt that remained after crystal growth.

## 1 Introduction

Considerable evidence points to the existence of dark matter that could represent 27% of the universe's total mass or energy [1-5]. One of the most stringent candidates for dark matter is the Weakly Interacting Massive Particles (WIMPs), which many experimental groups have extensively searched in the last few decades [6-13]. Despite attempting to find dark matter particles in numerous experiments, only the DAMA collaboration has claimed the observation of a dark matter signal through an annual modulation signal observed in the low-energy signal region [8,14-16]. However, there have been long-standing questions about this claim because no other experimental searches have observed similar signals [17]. Besides, no convincing explanation of the signal's origin has been proposed, regardless of the exact nature of the signal's dark matter.

The COSINE-100 experiment has been operating at Yangyang underground laboratory in Korea with a total of 106 kg of low-background NaI(Tl) detectors during the last six years [6,7,18-22]. Although many exciting results were published, reaching an unambiguous assumption on the annual modulation signal of the DAMA experiment is far from the conclusion [23,24]. It is mainly due to the observed background rate in the COSINE-100 detectors, which is 2.5 times higher than the background of the DAMA detectors [19,25]. To take the challenge in world competitive searches for low-mass dark matter and reach a definite conclusion for the DAMA/LIBRA, we are preparing the COSINE-200 experiment as the next phase of the COSINE-100 [17,26]. The main goal of the COSINE-200 is to develop 200 kg of ultra-low background NaI(Tl) crystals with a background level lower than those of the DAMA/LIBRA. To reach the physics goal of COSINE-200, we have been developing technology for the low-background NaI(Tl) detector that includes the mass production of ultra-low background NaI powder, crystal growing technique, and detector assembly [17,26,27]. The first step is preparing the ultra-low background NaI powder, in which the potassium concentration must be below 20 ppb and the lead concentration less than a few ppb. Radioactivity-wise, commercially available Astro-grade NaI powders from Sigma-Aldrich are suitable for ultra-low background NaI(Tl) crystal synthesis [17,28]. Still, their extremely high-cost demands independent development of mass purification technology. We have investigated a recrystallization technique to purify the NaI powder at a reasonable price [29]. The lab-scale procedure provided a satisfactory performance of the potassium and lead reduction. Based on successful lab-scale experiments, the mass purification facility was established at the Institute for Basic Science (IBS) in Daejeon, Korea [27]. For the last two years, we optimized operational parameters for the mass production of ultra-low background NaI powder. The yield efficiencies for the chemical process were balanced versus the products' purity. The processing conditions were adapted to recycle the mother solution and recover NaI from the melt residual after the crystal growth. Using developed technology, we have produced about 480 kg of the low-background powder with a production capability of 35 kg per two weeks. Using the purified NaI powder, radioactive background was reduced at least twice in a small size of NaI(Tl) crystal relative to the COSINE-100 crystals [30]. In this report, we summarize our experience, describe the mass purification facility, optimized raw powder purification, and the recovery of NaI from the mother solution and residual melt.

## 2   Materials and Methods

We use NaI powder from Merck (99.99(5)% purity, Optipure®) as an initial material. The potassium contamination in the specially ordered powder is below one ppm. High resistance, 18.2 MΩ·cm deionized (DI) water is a solvent to dissolve the NaI powder. We use absolute ethanol (~200 proof, HPLC grade, ACS) from the Scharlau to wash the recrystallized NaI crystals. Hydrophilic PTFE membrane filters with 1.0 $\mu$m pore size from the Advantec are used to separate the recrystallized NaI crystals from the mother liquor.

The mass production facility of the ultra-low background NaI powder is shown in Fig. 1 A. It consists of two main reactors (Fig. 1 B and C), a Nutsche filter unit (Fig. 1 D), two receivers (Fig. 1 E), and a conical dryer (Fig. 1 F). Operation of the whole system, including temperature control through the oil circulation system, is performed by the main controller in Fig. 1 G. The feed tank (Fig. 1 B) is used for the powder dissolution and pre-processing to prevent oxidation of iodide ions. Two main reactors in Fig. 1 B and C are connected, utilizing the polypropylene (PP) pipes that transfer the NaI solution from the feed tank to the mixing tank (Fig. 1 C), as shown in Fig. 1 A. A cartridge filter is installed in the middle of the PP pipelines to remove the insoluble impurities from the solution. The mixing tank performs the recrystallization using the temperature dependence of the NaI solubility in the water [31]. We evaporate water from the NaI solution until it becomes oversaturated at 110℃ (Fig. 2 A), then cool the mixing tank down to 30℃ while stirring the solution (Fig. 2 B). In this process, pure NaI crystals grow without agglomeration, while soluble impurities remain in the mother solution. The crystals are separated from the mother solution by the PTFE membrane filter (Fig. 2 C). The crystals are washed with chilled ethanol to rinse off the remaining mother liquor and impurities from the crystal surface.

The washed crystals are dried in the conical dryer (Fig. 2 D). The produced powders are packed in HDPE bottles and stored in the desiccators to avoid moisture absorption. The details of the facility and recrystallization procedure are described elsewhere in [27].

Radiopurity in the raw and purified powders and the mother solution from the purification process is measured by an inductively coupled plasma mass spectrometry (ICP-MS) and high-purity Germanium (HPGe) detector [32]. The water content in the produced powders is measured by the Karl-Fisher titrator.

## 3    Results

### 3.1    Raw powder purification

The main goal of our purification is to reduce internal potassium (K) contamination to less than 20 ppb. Tables 1 and 2 show the representative measurements from the raw powder purification process by ICS-MS and HPGe, respectively. As shown in Table 1, most of the potassium contamination coming from the raw powder was filtrated and concentrated in the mother solution. Potassium and lead concentrations in the purified powders were reduced by 20 and 80 times, resulting in final amounts of 11 ppb and 0.5 ppb, respectively. Significant reduction of Sr and Ba below the ppb level may indicate a reduction of radium, which belongs to the same family group of the periodic table. With a single crystallization procedure with about 40% yield efficiency, the purity of produced powder became similar to the Astro-grade powder. The impurities concentration in the mother solution were increased approximately twice as in the raw powder. Twenty days of HPGe counting using 1.2 kg of purified powder sampled in the Marinelli beaker reported only upper limits for $^{226}$Ra, $^{228}$Ac, $^{228}$Th, and $^{40}$K, as seen in Table 2.

To improve production capacity keeping the high quality of the product, we continually performed the raw powder purification with slightly different initial charges and recovery yields, as summarized in Table 3. Although the powder charge was increased from 40 kg to 64 kg, the purified product had similar purities from batch to batch. However, a high recovery yield of 58% provided considerable contamination of K, about 38 ppb. In case of the recovery yields were less than 50%, the purified powder contained consistently low contamination, especially K, about 10 ppb. To keep the consistent and high quality of the product, we ascertained a 50% yield efficiency at maximum for our purification process. Routine purification works have made our experience proficient for the last two years. Compared to the initial investigation shown in Ref. [27], we obtained consistently stable products with the required background level using the same purification facility. With the above-optimized purification parameters, the process took two weeks. Recrystallizing the raw powder took about three working days with 70 kg of the initial charge, and another seven working days were required to dry the wet crystals. With 40~50% recovery efficiency, 30~35 kg of purified powder could be produced in a cycle.

### 3.2    Mother solution recovery

After the purification process, the mother solution is the remaining product that is concentrated impurities from the initial material. In the optimized purification process, 50% of the initial charge was collected as the purified dry product. Another 35% of NaI remained in the mother solution, and 15% was washed out with ethanol, as shown in Fig. 3. In three cycles of the raw powder purification, the amount of NaI collected as the mother solution was enough for further recycling. We recovered this mother solution in the same manner but reduced the recovery efficiency from 50% to 35% due to the relatively high impurity level in the mother solution. As summarized in Table 4, the recovered crystals from the mother solution contained higher impurities than those obtained from the raw powder purification. The K contaminations varied from 18 to 50 ppb, proportional to the initial impurities in the mother solution. When the K content in the initial mother solution was higher than 1000 ppb, reaching the required 20 ppb of K was challenging with a single treatment. In this case, an additional recrystallization cycle of the powder was necessary to reach our goal of purity. However, following recrystallization of the crystals recovered from first mother solution (MS-1) was inefficient in production rate, so the rational K level in the initial solution must be lower than one ppm.

As shown in Fig. 3, after the separation of crystals from the MS-1, the second mother solution (MS-2) contained about 50% of NaI and accumulated most of the impurities. The MS-2 mostly had K content over one ppm. Double recrystallization would be unavoidable to recover this NaI remained. We did not consider recycling the MS-2 due to low recovery efficiency compared to the workforce required.

### 3.3 Residual melt recovery

We designed a large-size Kyropoulos grower to synthesize 120 kg NaI(Tl) crystal ingot [17]. In this grower, about 200 kg of NaI powder was loaded and melted in the quartz crucible. Crystal-growing trials using Merck raw powders were performed a couple of times with partial success. After pulling out the crystal ingot, many residues remained in the quartz crucible. Typical impurities in this melt were approximately twice higher as in the loaded powder due to the segregation effect. Nevertheless, the recovery of the residual melts was successfully made by achieving satisfied purity levels, as summarized in Table 5. The K concentration in the produced powders varied from 8 to 11 ppb. The purity of recovered NaI from the melt is expected to be much pure if we use the purified powder for mass crystal growth.

The process of recovering NaI from the collected residual melt differed from the original purification method because the melt contained a significant amount of insoluble quartz particles and dust. Before the usual operation, the NaI melt dissolved in water was filtered with the PTFE membrane filter. Considering the evaporation of iodine during the crystal-growing process, a three times higher dose of hydrogen iodide (HI) was introduced to reach pH 3.5.

### 3.4 Water content measurement

Sodium iodide is highly hygroscopic, and its chemical interaction with moisture produces NaOH when heated and causes corrosion of the quartz crucible used in the growing crystal [33]. Keeping the water content below 1000 ppm in the produced powder was crucial. All recrystallized powders were dried in two-step processes. In the first step, the wet powder was dried at 65℃ to avoid agglomerating the NaI powders with water inside the dryer. Then the temperature was increased to 130℃ to dry powder completely. The vapor released from the drying process was extracted with a vacuum pump. Initially, we used a chemical resistance air pump with relatively low pressure to protect the pump from corrosive vapor. As seen in Fig. 4, reaching moisture content below 1000 ppm in the dried powders with the previous set-up was impossible. We improved our drying system by introducing a high-pressure rotary pump with traps for corrosive vapor during the high-temperature drying process. We achieved the water content to less than 1000 ppm with modified set-up.

### 4 Discussion

A facility for mass production of the ultra-pure NaI powder for the COSINE-200 is well-operating with extensive parameter optimization. The purification of raw NaI powder, the recycling of the mother solution, and the recovery of NaI from the residual melt were performed in parallel. We have produced about 480 kg of low-background powder with a successful reduction of the internal contamination that is pure enough for the COSINE-200 detectors. The optimized parameters with a stable operation process have provided a maximum 35 kg powder production capacity in two weeks, but there is still room for improvement. If we increase the volume of the dryer 1.5 times, then two purification cycles can be performed in two weeks, increasing production capacity up to 70 kg. With improved capacity, successive double crystallization can help to reach a potassium level much lower than 5 ppb using the above-described facility. We can smoothly provide the ultra-low background NaI powder for the mass production of the NaI(Tl) crystals for the COSINE-200 experiment.

## 5 Acknowledgments

This work is supported by the Institute for Basic Science (IBS) under project code IBS-R016-A1.

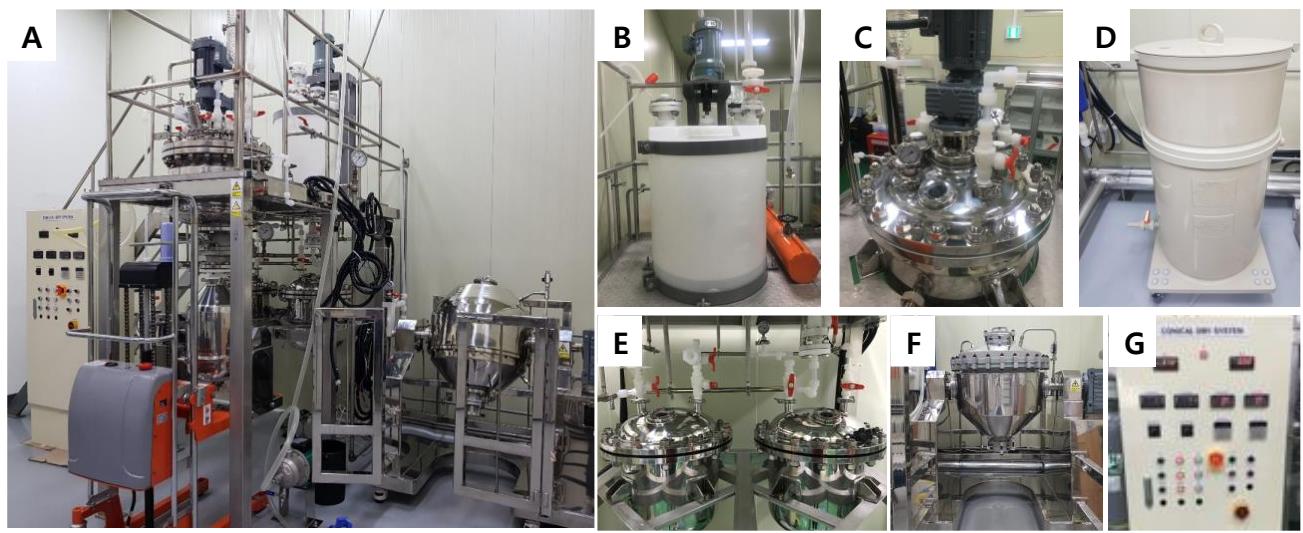

**Figure 1.** (A) Mass purification facility, (B) Feed tank for dissolving the NaI, (C) Mixing tank for boiling solution and recrystallization process, (D) Filter unit for separation of NaI crystal and mother liquor, (E) Receiver tanks for collecting vapor from mixing tank and dryer, (F) Conical dryer for the NaI powder drying, (G) Main controller to control all the equipment.

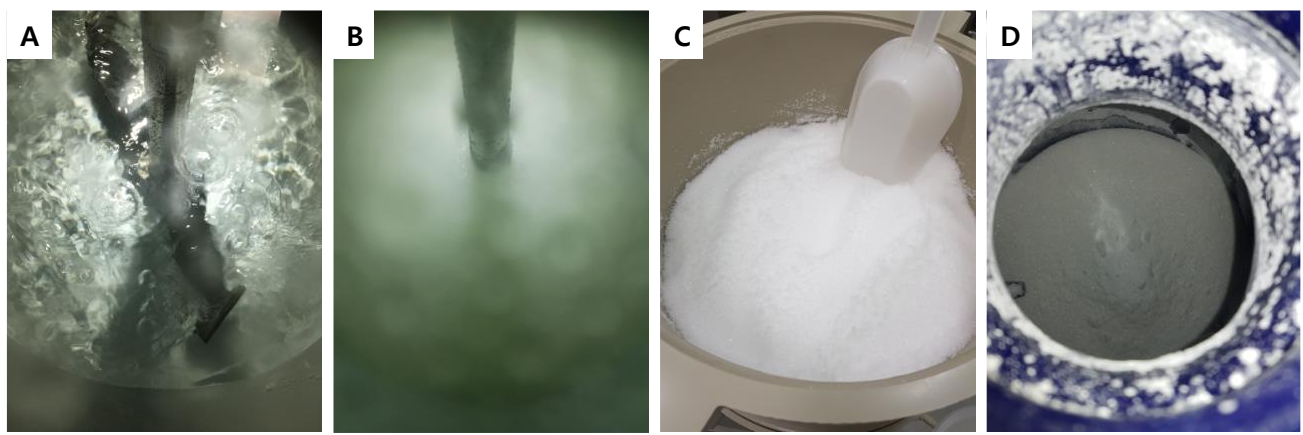

**Figure 2.** (A) Boiling of solution with stirring, (B) Recrystallized NaI crystal with mother liquor, (C) Filtrated and washed NaI crystal on the filter unit, (D) Dried NaI powder in the conical dryer.

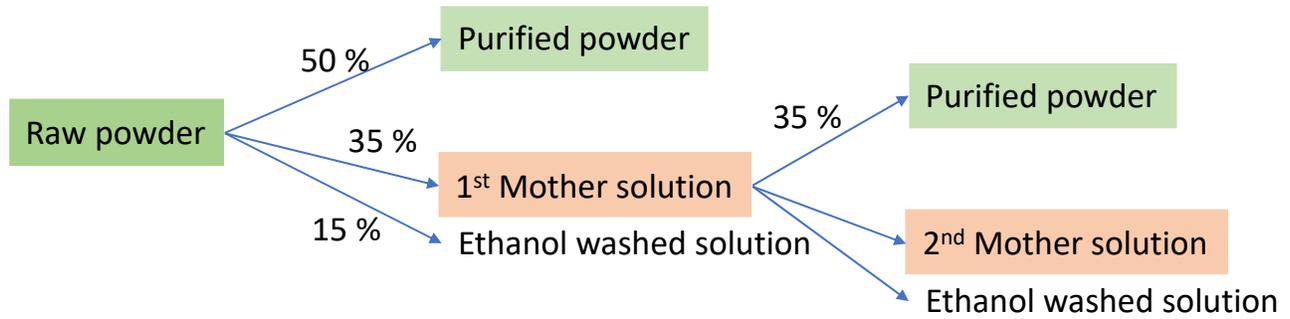

**Figure 3.** Material balances in the NaI recovery cycle.

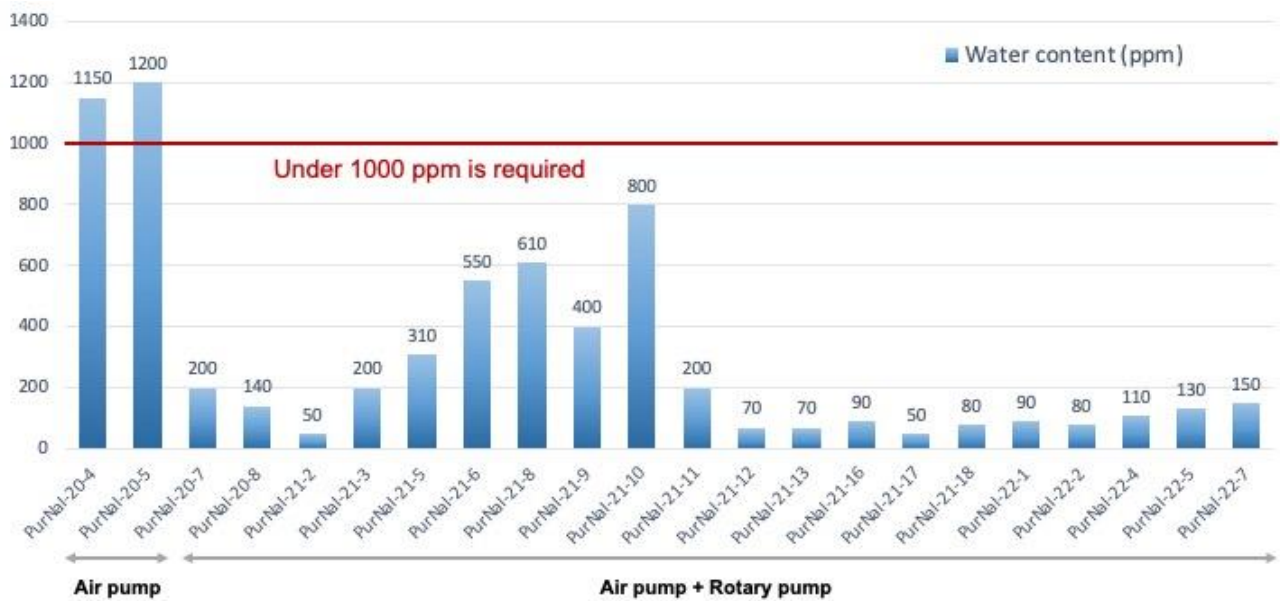

**Figure 4.** The water content measurement results by Karl-Fisher titrator for different batches of produced powders

**Table 1.** Representative ICP-MS results of raw and purified powders vs. Astro-grade powder's purity. Values are given at 90% C.L, and upper limits are given at 95% C.L.

| Description | K ppb | Fe ppb | Sr ppb | Ba ppb | Pb ppb | Th ppt | U ppt |
|---|---|---|---|---|---|---|---|
| Astro grade | 5±3 | 110±20 | 0.3±0.1 | 0.6±0.1 | 0.8±0.1 | < 6 | < 6 |
| Merck-raw powder | 250±90 | 33±6 | 19±1 | 3.0±0.4 | 40±2 | < 6 | < 6 |
| Purified powder (20-5) | 11±1 | < 10 | 0.3±0.1 | 0.9±0.1 | 0.5±0.1 | < 6 | < 6 |
| Mother solution (20-5) | 550±120 | < 200 | 38±2 | 9±1 | 60±4 | < 6 | < 6 |

**Table 2.** Representative HPGe result of purified powder from raw powder purification. The upper limits are given at 90% C.L.

| $^{226}$Ra ($^{238}$U) | $^{40}$K | $^{228}$Ac | $^{228}$Th |
|---|---|---|---|
| < 0.56 mBq/kg | < 5.64 mBq/kg | < 1.10 mBq/kg | < 0.71 mBq/kg |

**Table 3.** The ICP-MS results of purified powders in different batches of raw powder purification. Values are given at 90% C.L, and upper limits are given at 95% C.L.

| Sample No. | Initial charge | Recovery yield | K ppb | Fe ppb | Sr ppb | Ba ppb | Pb ppb | Th ppt | U ppt |
|---|---|---|---|---|---|---|---|---|---|
| 20-5 | 40 kg | 44% | 11±1 | < 10 | 0.3±0.1 | 0.9±0.1 | 0.5±0.1 | < 6 | < 5 |
| 20-7 | 50 kg | 41% | 10±1 | < 10 | 0.1±0.1 | 0.3±0.1 | < 0.3 | < 3 | < 5 |
| 20-8 | 50 kg | 39% | 6.4±0.1 | < 10 | 0.1±0.1 | 0.7±0.1 | < 0.3 | < 3 | < 5 |
| 21-5 | 53 kg | 42% | 5.4±0.3 | < 10 | 0.2±0.1 | 0.4±0.1 | 0.5±0.1 | < 5 | < 3 |
| 21-8 | 60 kg | 58% | 38±2 | < 10 | 0.4±0.1 | 0.3±0.1 | 0.5±0.1 | < 7 | < 7 |
| 22-5 | 64 kg | 35% | 11±1 | < 7 | 0.4±0.1 | 1.6±0.2 | 0.9±0.5 | < 100 | < 20 |

**Table 4.** The ICP-MS result of the mother solution recovery experiment in different batches of mass production. It is marked as (M) for the naming. In this experiment, the Initial solution means the initial mother solution, and the Wet crystal means recrystallized and washed crystal. If the purity is not accepted, then additional recrystallization is required, so the purity was confirmed first by ICP-MS before drying and then dried thoroughly. The wet crystal consists of ~73% NaI and extra water and ethanol, so the impurity concentration is calculated as 73% NaI. Values are given at 90% C.L, and upper limits are given at 95% C.L.

| Sample No. | Material | K (ppb) | Fe (ppb) | Sr (ppb) | Ba (ppb) | Pb (ppb) | Th (ppt) | U (ppt) |
|---|---|---|---|---|---|---|---|---|
| 21-4(M) | Initial sol. | 330±40 | N/A | 20±1 | 6.3±0.2 | 41±4 | < 5 | < 3 |
|  | Wet cryst. | < 40 | N/A | 0.4±0.1 | 0.1±0.1 | 1.2±0.1 | < 5 | < 3 |
| 21-7(M) | Initial sol. | 470±10 | N/A | 34±1 | 7.3±0.2 | 56±1 | < 7 | < 7 |
|  | Wet cryst. | < 50 | N/A | 1.2±0.1 | 0.2±0.1 | 2.0±0.1 | < 7 | < 7 |
| 21-11(M) | Initial sol. | 610±30 | 16±1 | 40±2 | 10±1 | 88±12 | < 7 | < 7 |
|  | Wet cryst. | 18±1 | < 7 | 1.0±0.1 | 0.2±0.1 | 4.5±0.3 | < 7 | < 7 |
| 22-2(M) | Initial sol. | 1010±150 | 8±1 | 16±1 | 15±1 | 86±4 | < 4 | < 4 |
|  | Dry powder | 21±2 | < 7 | 0.2±0.1 | 0.7±0.1 | 1.0±0.1 | < 4 | < 4 |
| 20-3(M) | Initial sol. | 1170±120 | 39±2 | 33±2 | 12±1 | 60±2 | < 6 | < 5 |
|  | Dry powder | 44±5 | 14±1 | 1.0±0.1 | 0.4±0.1 | 2.0±0.1 | < 6 | < 5 |

**Table 5.** The ICP-MS result of residual melt recovery experiment in different batches of mass production. It is marked as (RM) for the naming, and the Initial solution is the residual melt solution after dissolving melt and filtration of the quartz particles. The Wet crystal samples were taken after recrystallization and washing with ethanol. Values are given at 90% C.L, and upper limits are given at 95% C.L.

| Sample No. | Material | K (ppb) | Fe (ppb) | Sr (ppb) | Ba (ppb) | Pb (ppb) | Th (ppt) | U (ppt) |
|---|---|---|---|---|---|---|---|---|
| 21-12(RM) | Initial sol. | 730±10 | 20±2 | 10±1 | 8.0±0.6 | 143±12 | < 7 | < 7 |
|  | Wet cryst. | 8±1 | < 10 | 0.4±0.1 | 0.3±0.1 | 5±1 | < 7 | < 7 |
| 21-13(RM) | Initial sol. | 540±20 | N/A | 10±1 | 5.1±0.2 | 95±10 | < 7 | < 7 |
|  | Wet cryst. | < 50 | N/A | 0.1±0.1 | < 0.1 | < 0.3 | < 7 | < 7 |
| 22-1(RM) | Initial sol. | 390±10 | N/A | 8±1 | 6.4±0.2 | 40±4 | < 4 | < 4 |
|  | Dry powder | 8±1 | < 7 | 0.1±0.1 | 0.3±0.1 | 0.7±0.1 | < 4 | < 4 |
| 22-4(RM) | Initial sol. | 570±10 | N/A | 15±2 | 6.7±0.5 | 5±1 | < 4 | < 4 |
|  | Dry powder | 11±4 | < 7 | 0.1±0.1 | 0.3±0.1 | < 0.3 | < 4 | < 4 |